\documentclass[amsmath,amssymb,aps,prb,reprint]{revtex4-2}

\usepackage{graphicx}
\usepackage{pifont}
\usepackage{bm}
\usepackage{xcolor}
\usepackage[colorlinks=true,linkcolor=blue,anchorcolor=blue,citecolor=blue,urlcolor=blue]{hyperref}

\begin{document}

\preprint{APS/123-QED}

\title{Exotic Topological Phenomena in Chiral Superconducting States on Doped Quantum Spin Hall Insulators with Honeycomb Lattices}

\author{Junkang Huang$^{1,2}$}
\author{Tao Zhou$^{1,2,3}$}
\email{tzhou@scnu.edu.cn}
\author{Z. D. Wang$^3$}
\email{zwang@hku.hk}

\affiliation{$^1$Guangdong Provincial Key Laboratory of Quantum Engineering and Quantum Materials, School of Physics, South China Normal University, Guangzhou 510006, China\\
	$^2$Guangdong-Hong Kong Joint Laboratory of Quantum Matter, Frontier Research Institute for Physics, South China Normal University, Guangzhou 510006, China\\
$^3$Guangdong-Hong Kong Joint Laboratory of Quantum Matter, Department of Physics, and HK Institute of Quantum Science $\&$ Technology, The University of Hong Kong, Pokfulam Road, Hong Kong, China
}

\date{\today}

\begin{abstract}
We have conducted a theoretical investigation of the topological phenomena associated with chiral superconducting pairing states induced in a doped Kane-Mele model on a honeycomb lattice. Through numerical analysis, we have obtained exotic phase diagrams for both the $d+id$ and $p+ip$ superconducting states. In the case of the $d+id$ pairing state, higher Chern number states with $\left| C \right| = \pm 4$ emerge. The Chern number decreases as the spin-orbit coupling is introduced. For the $p+ip$ pairing state, additional phase transition lines are present in the overdoped region near the Van Hove singularity point, leading to the emergence of higher Chern number phases with $\left| C \right| = \pm 6$. These higher Chern number phases are further verified through the bulk-edge correspondence.
To understand the origin of the exotic topological phase diagrams in the chiral superconducting state, we have examined the electronic structure at the phase transition lines. This investigation provides insight into the complex interplay between chiral superconductivity and topological properties, potentially paving the way for the discovery of new materials with unique topological properties.

\end{abstract}
\maketitle

\section{\label{Intro}Introduction}
Chiral topological superconductors have been both theoretically and experimentally investigated, drawing significant research interest due to their fascinating properties~\cite{IOP1361-648X/abaa06,IOP0034-4885/79/5/054502}. The chiral pairing term breaks time-reversal symmetry, leading to chiral Majorana zero modes and edge currents with promising potential for quantum computation. Various chiral superconducting pairing states have been proposed in numerous systems~\cite{science.1248552,s41567-020-0822-z,science.abb0272,s41586-020-2122-2,s41467-023-38688-y,PhysRevLett.102.117007,PhysRevB.82.174511,PhysRevB.101.100506,s41586-019-1596-2,PhysRevB.87.180503,PhysRevB.89.020509,PhysRevMaterials.3.104802,PhysRevB.102.020503,PhysRevB.91.140506,JPSJ.85.033704,PhysRevB.102.134511,PhysRevLett.115.267001}.

Honeycomb lattice systems are excellent candidates for achieving chiral superconductivity. One well-known material system with honeycomb lattice symmetry is graphene-based materials. Superconductivity in this family of materials has previously attracted considerable research interest. Various theoretical techniques, such as symmetry analysis, exact numerical study, random phase approximation, renormalization group method, dynamic cluster approximation, kinetic-energy-driven superconductivity, and quantum Monte Carlo method, have extensively suggested that the superconducting pairing symmetry in graphene-based materials could be either $p+ip$ pairing symmetry or $d+id$ pairing symmetry~\cite{PhysRevB.84.121410,PhysRevB.86.020507,PhysRevB.89.144501,PhysRevB.90.245114,PhysRevB.92.085121,Nandkishore2012,Xiao_2016,HUANG2019310,PhysRevB.101.155413,PhysRevB.94.115105,PhysRevB.102.125125,PhysRevB.107.245106,lan2023doping}. 
Chiral superconductivity may serve as a useful platform for exploring nontrivial topology in the honeycomb lattice system~\cite{PhysRevB.108.134515}.

Experimentally, superconductivity signatures have been observed in various graphene-based materials~\cite{Xue2012,Ichinokura2016,doi:10.1073/pnas.1510435112,Chapman2016,Cao2018,doi:10.1126/science.abm8386}.
Besides graphene-based materials, superconductivity has also been discovered in $\beta$-MNCl (M = Hf, Zr)~\cite{https://doi.org/10.1002/adma.19960080917,Yamanaka1998}, SrPtAs~\cite{doi:10.1143/JPSJ.80.055002,PhysRevB.87.180503}, silicene-based materials~\cite{WOS:000315597000027,Zhang2015,PhysRevLett.111.066804}, hydrogenated germanene~\cite{Xi2022}, and stanene~\cite{Liao2018,PhysRevLett.128.206802}. These materials also exhibit honeycomb lattice symmetry.
On the other hand, it has been reported that the honeycomb lattice and unconventional superfluidity can be realized in cold atom systems~\cite{Soltan-Panahi2012,Tarruell2012}, offering another platform for exploring chiral pairing states and potential exotic topological behaviors.

Another non-trivial topological state that has garnered significant attention is the quantum spin Hall insulating (QSHI) state. The QSHI state in a honeycomb lattice can be characterized by the Kane-Mele (KM) model, which was originally proposed to describe the topological properties of graphene~\cite{PhysRevLett.95.226801}. It has subsequently been proposed for implementation in materials such as silicene, germanene, stanene, $\mathrm{Pt}_2\mathrm{HgSe}_3$, and $\mathrm{Pd}_2\mathrm{HgSe}_3$ \cite{JPSJ.84.121003,acs.nanolett.9b02689,nmat4802,PhysRevB.84.195430,PhysRevLett.109.055502,PhysRevLett.120.117701,PhysRevMaterials.3.074202}. In the QSHI state, energy bands are fully gapped, rendering the material a topological insulator. The Fermi level of a material can be artificially adjusted through gate voltage or chemical doping. As a result, the metallic state is realized when the Fermi level intersects the energy bands. At low temperatures, the superconducting state should be the ground state, with an additional effective pairing interaction. It has previously been reported that the Fermi level of monolayer QSHI material WTe$_2$ can be gate-tuned, subsequently inducing superconductivity~\cite{doi:10.1126/science.aar4426,doi:10.1126/science.aar4642}. For graphene, it has also been reported that the material can be heavily overdoped, with the Fermi level being tuned beyond the Van Hove Singularity (VHS) point~\cite{PhysRevLett.125.176403}. Consequently, unconventional superconductivity may be realized due to fermiology near the VHS point~\cite{PhysRevB.103.174513}. Hence, investigating the topological behavior in the superconducting state of doped QSHI material on a honeycomb lattice is fundamentally fascinating, as the interplay between distinct non-trivial topologies could result in a considerably more complex phase diagram and even more exotic topological phenomena.

In this paper, motivated by the considerations above, we investigate the topological properties of a honeycomb lattice system by introducing chiral superconductivity into the doped KM model. The topological properties can be characterized by defining the Chern number of the system. We obtain a rich phase diagram numerically, demonstrating the existence of high Chern number phases. The presence of higher Chern number states strongly depends on lattice symmetry and can be investigated by examining the Dirac cones around phase transition lines. These higher Chern number states can be further confirmed by exploring the edge states of the system. The existence of multiple Majorana zero modes at the system edges may have potential applications in topological quantum computation \cite{s41567-019-0517-5,pnas.1810003115}. Experimentally, the Chern number of a chiral topological system relates to the edge current or the anomalous hall conductance~\cite{PhysRevB.91.094507,PhysRevB.90.224519,PhysRevB.107.224517}. However, its relationship still requires further investigation. The rich and exotic phase diagram presented here offers a new platform to explore the relationship between the Chern number and observable physical quantities in a chiral topological superconducting system.

The organization of this paper is as follows: In Section II, we describe the model and elaborate on the formalism. In Section III, we present the numerical results, including phase diagrams, energy spectra, and corresponding boundary spectral functions for different phases. In Section IV, we discuss the origins of the numerical results. Finally, we provide a brief summary of our work in Section V.

\section{\label{sec:Model}Model and Formalism}
We begin by defining the Hamiltonian on a honeycomb lattice, which includes both the normal state component and the chiral superconducting pairing component. The complete Hamiltonian can be represented in momentum space as $H = \sum_{\bf k} \Psi^{\dagger}({\bf k}) H({\bf k}) \Psi({\bf k})$, where $H({\bf k})$ is expressed in an $8\times 8$ matrix,

\begin{eqnarray}
H\left(\bf{k}\right) = \left( \begin{array}{cccc}
{H_{0}\left(\bf{k}\right)}&{\Delta_{D/P}\left(\bf{k}\right)}\\
{\Delta^{\dagger}_{D/P}\left(\bf{k}\right)}&{-H_{0}^*\left(\bf{k}\right)}
\end{array} \right),
\end{eqnarray}
The basis vector $\Psi^{\dagger}\left(\bf{k}\right)$ is expressed as $( C_{\bf{k}A\uparrow}^{\dagger}, C_{\bf{k}A\downarrow}^{\dagger}, C_{\bf{k}B\uparrow}^{\dagger}, C_{\bf{k}B\downarrow}^{\dagger}, $ $C_{\bf{k}A\uparrow}, C_{\bf{k}A\downarrow}, C_{\bf{k}B\uparrow}, C_{\bf{k}B\downarrow} )^T$. $H_{0}\left(\bf{k}\right)$ is the $4\times 4$ normal states Hamiltonian matrix, taken as the doped KM model \cite{PhysRevLett.95.226801}, expressed as, 
\begin{eqnarray}
H_{0}\left(\bf{k}\right) = \left( \begin{array}{cccc}
{h_{SO}\left(\bf{k}\right)}&{h_t\left(\bf{k}\right) + h_{R}\left(\bf{k}\right)}\\
{\left[ h_t\left(\bf{k}\right) + h_{R}\left(\bf{k}\right) \right]^{\dagger}}&{-h_{SO}\left(\bf{k}\right)}
\end{array} \right) - \mu,
\end{eqnarray}
where $h_t\left(\bf{k}\right)$, $h_{SO}\left(\bf{k}\right)$ and $h_R\left(\bf{k}\right)$ are the nearest neighbor hopping term, spin-orbit coupling term and Rashba term, respectively.
\begin{eqnarray}
h_t\left(\bf{k}\right) &=& -ts_0 \sum_{i} \exp\left( i\bf{k} \cdot \bf{a_i} \right), \\
h_{SO}\left(\bf{k}\right) &=& -2\lambda_{SO} s_3 \sum_{i} \sin\left( \bf{k} \cdot \bf{b_i} \right), \\
h_R\left(\bf{k}\right) &=& i\lambda_R \sum_i \exp\left( i\bf{k} \cdot \bf{a_i} \right) \left( \bf{s}\times \frac{\bf{a_i}}{\left| \bf{a_i} \right|} \right)_z,
\end{eqnarray}
where $t$ is the nearest-neighbor hopping constant. $s_0$ and $s_i$ $(i=1,2,3)$ are the identity matrix and the Pauli matrices in the spin channel, respectively. Vectors ${\bf a_i}$ and ${\bf b_i}$ are expressed as: ${\bf a_1} = (\sqrt{3}a/3,0)$, ${\bf a_2} = (-\sqrt{3}a/6,a/2)$, ${\bf a_3} = (-\sqrt{3}a/6,-a/2)$, and ${\bf b_{1}} = (0,-a)$, ${\bf b_{2}} = (\sqrt{3}a/2,a/2)$, ${\bf b_{3}} = (-\sqrt{3}a/2,a/2)$.

$\Delta_{D/P}\left(\bf{k}\right)$ is the $d+id / p+ip$ pairing term, written as,
\begin{eqnarray}
\Delta_{D}\left(\bf{k}\right) &=& i\Delta_0^D \sigma_1 s_2 ( e^{i \sigma \bf{k} \cdot {\bf a_1}} + e^{i4\pi/3} \cdot e^{i \sigma \bf{k} \cdot {\bf a_2}} \nonumber \\
&& + e^{i8\pi/3} \cdot e^{i \sigma \bf{k} \cdot {\bf a_3}} ), \\
\Delta_{P}\left(\bf{k}\right) &=& -\Delta_0^P \sigma_2 s_2 ( e^{i \sigma \bf{k} \cdot {\bf a_1}} + e^{i2\pi/3} \cdot e^{i \sigma \bf{k} \cdot {\bf a_2}} \nonumber \\
&&+ e^{i4\pi/3} \cdot e^{i \sigma \bf{k} \cdot {\bf a_3}} ),
\end{eqnarray}
where $\sigma_i$ are Pauli matrices in the sublattice channel. 
 $\sigma$ takes $+1$ or $-1$ depending on the pairing orientation: A to B or B to A, respectively.

$\Delta^{D/P}_0$ is the superconducting order parameter magnitude,
which can be calculated self-consistently in the momentum space as,
\begin{eqnarray}
	\label{eq:self}
	\Delta_{0}^D &=& \frac{V}{2} \sum_{{\bf k}n} u^*_{1n\bf{k}}u_{8n\bf{k}} \tanh \left( \frac{\beta E_{n_{\bf k}}}{2} \right) ( e^{i \bf{k} \cdot {\bf a_1}} + \nonumber \\ && e^{i4\pi/3} \cdot e^{i \bf{k} \cdot {\bf a_2}} 
	 + e^{i8\pi/3} \cdot e^{i \bf{k} \cdot {\bf a_3}} ), \\
	\Delta_{0}^P &=& \frac{V}{2} \sum_{{\bf k}n} u^*_{1n\bf{k}}u_{8n\bf{k}} \tanh \left( \frac{\beta E_{n_{\bf k}}}{2} \right) (e^{i \bf{k} \cdot {\bf a_1}} + \nonumber \\
	&& e^{i2\pi/3} \cdot e^{i \bf{k} \cdot {\bf a_2}} + e^{i4\pi/3} \cdot e^{i \bf{k} \cdot {\bf a_3}} ),
\end{eqnarray}
with $V$ being the attractive pairing interaction. 
 $u_{in\bf{k}}$ is the $i$-th component of the vector $u_{n\bf{k}}$.
$u_{n\bf{k}}$ and $E_{n_{\bf k}}$ are the $n$-th eigenvectors and $E_{n_{\bf k}}$ are eigenvectors and eigenvalues of the Hamiltonian matrix $H({\bf k})$.

When the chiral superconducting pairing term is present, the time reversal symmetry of the Hamiltonian is broken. As a result, the Chern number can be utilized to describe the nontrivial topology of the system, express as \cite{jpsj.74.1674},
\begin{eqnarray}
\label{EQ:chern}
C &=& \frac{1}{2\pi i} \sum_{\bf k} \tilde{F}_{xy} \left({\bf {k}}\right).
\end{eqnarray}
where $\tilde{F}_{xy}$ is the lattice field strength that can be defined by 
\begin{eqnarray}
\tilde{F}_{xy} \left({\bf {k}}\right) \equiv \ln \frac{ U_x\left({\bf k}\right) U_y\left({\bf k} + {\hat{x}}\right) }{U_x\left({\bf k} + {\hat{y}}\right) U_y \left({\bf k}\right)},
\end{eqnarray}
with
\begin{eqnarray}
U_{\alpha}\left({\bf k}\right) = \frac{\det \Phi^{\dagger}\left({\bf k}\right) \Phi\left({\bf k} + {\hat{\alpha}} \right)}{\left| \det \Phi^{\dagger}\left({\bf k}\right) \Phi\left({\bf k} + {\hat{\alpha}} \right) \right|},
\end{eqnarray}
where $\Phi({\bf k})$ is an $8\times 4$ matrix, which is constructed
by column-wise packing the four occupied eigenstates with $\Phi({\bf k})=(u_{1\bf{k}},u_{2\bf{k}},u_{3\bf{k}},u_{4\bf{k}})$.

To investigate the edge states, we apply a periodic boundary condition in the $x$ direction and a zigzag boundary condition in the $y$ direction $(1\leq y \leq N_y)$. The Hamiltonian with a partial open boundary condition can be expressed as $H = \sum_{j,{k_x}} \Psi^{\dagger}\left( j,{k_x} \right) \hat{M}\left( {k_x} \right) \Psi\left( j,{k_x} \right)$, where $\hat{M}( {k_x} )$ is an $8N_y\times 8N_y$ matrix and $\Psi_{j,{k_x}}^{\dagger}$
is the basis vector with
$\Psi_{j,{k_x}}^{\dagger} = ( C_{j,{k_x}A\uparrow}^{\dagger}, C_{j,{k_x}A\downarrow}^{\dagger}, C_{j,{k_x}B\uparrow}^{\dagger}, C_{j,{k_x}B\downarrow}^{\dagger}, C_{j,{k_x}A\uparrow},$ $C_{j,{k_x}A\downarrow}, C_{j,{k_x}B\uparrow}, C_{j,{k_x}B\downarrow})$ $(j=1,2,\cdots,N_y)$.

Then the spectral function can be calculated by,
\begin{eqnarray}
A_j\left( {k_x},\omega \right) = -\frac{1}{\pi} \sum_{p = 1}^4\sum_n  \mathrm{Im}\frac{\mid u_{m+p,n}(k_x)\mid^2}{\omega - E_n({k_x}) + i\Gamma},
\end{eqnarray}
with $m=4(j-1)$. $u_{m+p,n}(k_x)$ and $E_n({k_x})$ are eigenvectors and eigenvalues of the matrix $\hat{M}( {k_x} )$, respectively.

In the following presented results, without loss of generality, we use the hopping constant $t$ and the lattice constant $a$ as the energy unit $(t=1)$ and the length unit $(a=1)$, respectively. Other parameters are taken as $\lambda_R = 0.05$ and $\Gamma=0.01$.

\begin{figure}[tp]
\centering
\includegraphics[width = 8cm]{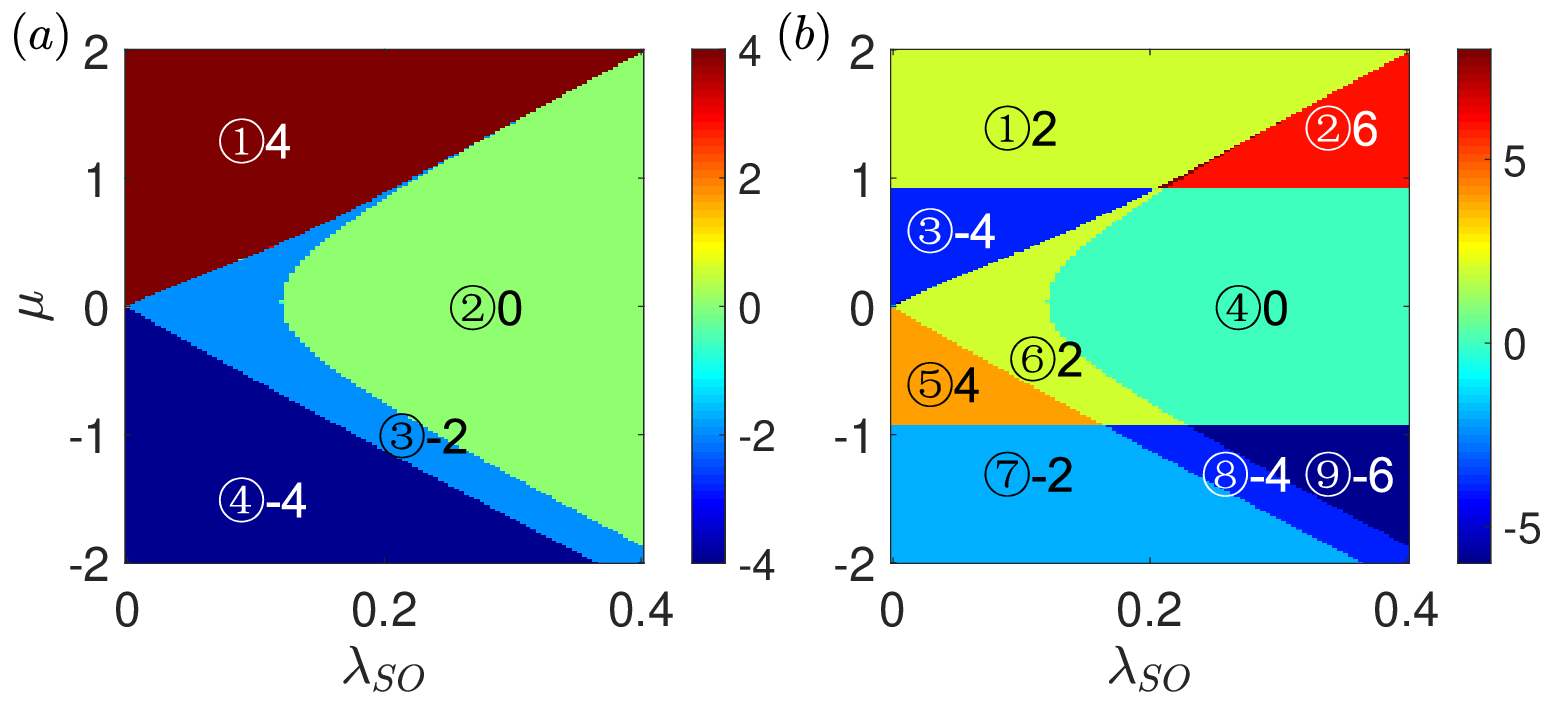}
\caption{\label{fig:PD} Topological phase diagrams depicting the dependence of Chern number $C$ on parameter sets $(\mu,\lambda_{SO})$ in distinct pairing states with the gap magnitude $\Delta_0=0.2$: (a) $d+id$ pairing state and (b) $p+ip$ pairing State.
}
\end{figure}

\section{\label{sec:Results}Results}
In the chiral superconducting state, the topological properties are described by the Chern number [Eq.(8)]. The Chern number should, in principle, depend strongly on the chemical potential $\mu$ and the spin-orbit coupling strength $\lambda_{SO}$. We present the $(\lambda_{SO},\mu)$-dependent topological phase diagram for the $d+id$ and $p+ip$ superconducting states in Figs.~\ref{fig:PD}(a) and \ref{fig:PD}(b), respectively. 

For the $d+id$ pairing symmetry, as seen in Fig.~~\ref{fig:PD}(a), the phase diagram includes four regions (indicated as \ding{172}$-$\ding{175}) with the Chern number $C$ equal to $4$, $0$, $-2$, and $-4$, respectively. For the $p+ip$ pairing symmetry, as seen in Fig.~\ref{fig:PD}(b), the phase diagram comprises nine regions (\ding{172}$-$\ding{180}). The phase diagram in the regime $-0.9<\mu<0.9$ is qualitatively consistent with that of the $d+id$ pairing symmetry. However, additional phase transition lines exist at the chemical potentials around $\mu=\pm 0.9$. Consequently, the phase diagram for the heavily overdoped regions with $|\mu|>0.9$ is significantly different, and higher Chern number phases with $C=6$ and $C=-6$ emerge.

Due to the existence of the Rashba term, the phase diagrams for both pairing symmetries are electron-hole asymmetric. The phase diagram for the electron-doped ($\mu>0$) region differs from that for the hole-doped ($\mu<0$) region. For the $d+id$ pairing symmetry, the $C=-2$ phase in the hole-doped region is significantly larger than in the electron-doped region. 

For the $p+ip$ pairing symmetry, in the heavily over hole-doped region ($\mu<-0.9$), there are three phases with $C=-2$, $-4$, and $-6$, as well as two phase transition lines. The Chern number decreases by 2 when crossing each transition line. In the heavily over electron-doped region ($\mu>0.9$), there are two stable topological phases with $C=2$ and $C=6$. Between these two phases lies a rather narrow region with $C=8$. This phase is unstable and exists only at several lattice sites. In fact, as the spin-orbit strength $\lambda_{SO}$ increases, there are still two phase transition lines. The Chern number increases by 6 when crossing the first line and decreases by 2 when crossing the second one.

\begin{figure}[tp]
\centering
\includegraphics[width = 8cm]{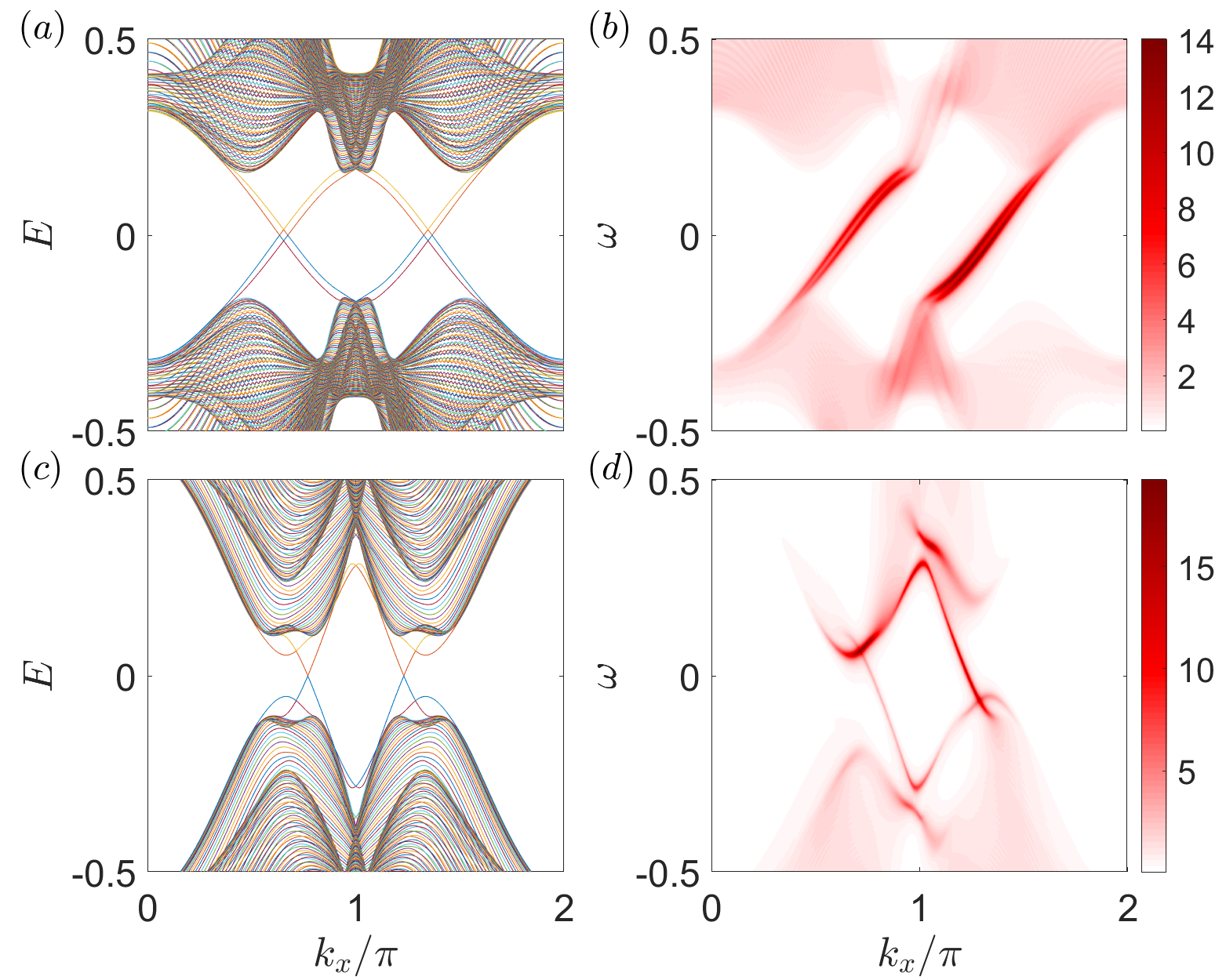}
\caption{\label{fig:d+id} Energy bands and spectral functions at the system edge $(y=1)$ in $d+id$ superconducting pairing states with zigzag boundary Condition along the $y$-Direction, varying with parameter sets $(\lambda_{SO},\mu)$. Panels (a) and (b) showcase the energy bands and spectral function in region \ding{172} $(C=4)$ for $(\lambda_{SO},\mu)=(0.1,1.1)$, respectively. Panels (c) and (d) display the energy bands and spectral function in region \ding{174} $(C=-2)$ for $(\lambda_{SO},\mu)=(0.15,-0.6)$, respectively.
}
\end{figure}

\begin{figure}
\centering
\includegraphics[width = 8cm]{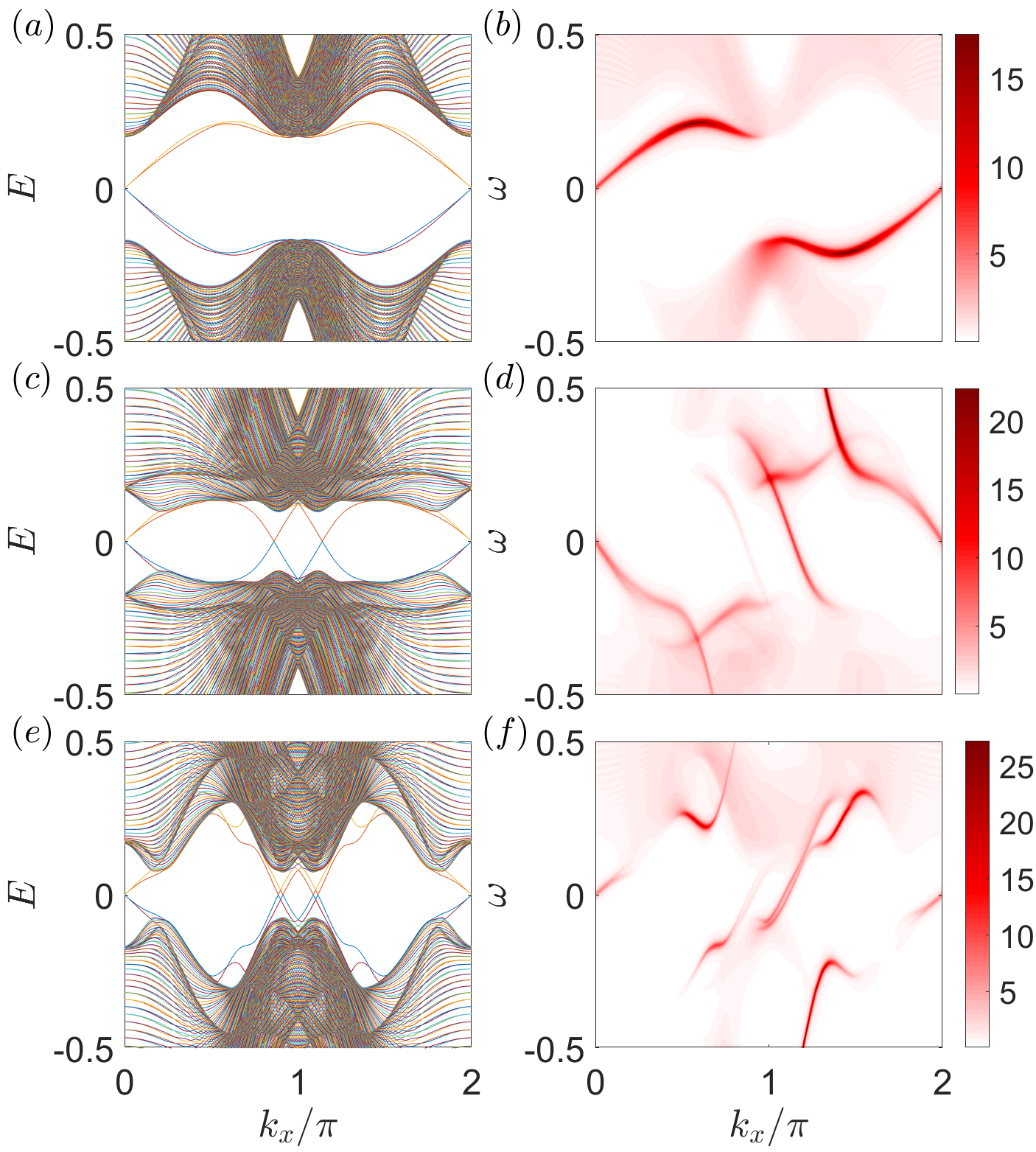}
\caption{\label{fig:p+ip} Similar to Fig.~\ref{fig:d+id}, but for numerical results in $p+ip$ superconducting states. Panels (a) and (b) display results in region \ding{172} (C=2) with $(\lambda_{SO},\mu)=(0.08,1.1)$. Panels (c) and (d) show results in region \ding{179} (C=-4) with $(\lambda_{SO},\mu)=(0.22,-1.1)$. Panels (e) and (f) depict results in region \ding{173} (C=6) with $(\lambda_{SO},\mu)=(0.3,1.1)$. }
\end{figure}

According to the bulk-edge correspondence, a topological nontrivial state with a Chern number $C=N$ should give rise to $2N$ edge states when considering the aforementioned partial open boundary condition, resulting in $N$ pairs of Majorana zero modes at the system edges. We now aim to verify the topological phase with different Chern numbers by investigating the edge states of the system while employing a zigzag boundary condition along the $y$-direction. The energy bands and site-dependent spectral functions can then be obtained through the diagonalization of the Hamiltonian matrix.
We present the numerical results of the energy bands and spectral functions at the system edge $(y=1)$ in various topological regions for the $d+id$ and $p+ip$ pairing symmetries in Figs.~\ref{fig:d+id} and \ref{fig:p+ip}, respectively. 

In the $d+id$ pairing state, there are three topologically nontrivial phases with Chern numbers of $4$, $-2$, and $-4$. We present the numerical results of the energy bands for the regions with $C=4$ and $C=-2$ in Figs.~\ref{fig:d+id}(a) and \ref{fig:d+id}(c), respectively. The corresponding spectral functions are shown in Figs.~\ref{fig:d+id}(b) and \ref{fig:d+id}(d), respectively. As can be seen, in the $C=4$ region, there are eight energy bands crossing the Fermi energy, resulting in four pairs of Majorana zero modes at the momenta $k_x=\pm 0.65\pi\pm 0.02\pi$. These energy bands originate from the edge states, and the edge modes are chiral, as observed from the spectral function spectra at the system edge [Fig.~\ref{fig:d+id}(b)]. The four energy bands with positive slopes (corresponding to the quasiparticles with positive velocity) are contributed by the $y=1$ edge. The other four energy bands with negative slopes are contributed by the $(y=N_y)$ edge (not presented here).

In the $C=-2$ region, similarly, there are four energy bands crossing the Fermi energy, corresponding to two pairs of Majorana zero modes at the momenta $\bm{k_x} = \pm 0.77\pi$. The Chern number is consistent with the number of edge states at the $y=1$ edge. The quasiparticle energy bands and spectral functions in the $C=-4$ region are qualitatively similar to those in the $C=4$ region; therefore, the numerical results are not presented here.


We now present the numerical results of the energy bands and spectral functions in the $p+ip$ superconducting state. The phase diagram in the $p+ip$ state is more complex, with topological phases having higher Chern numbers $C=\pm 6$ emerging. The numerical results of the energy bands and spectral functions with Chern numbers $2$, $-4$, and $6$ are displayed in Fig.~\ref{fig:p+ip}. As observed, the number of edge states is consistent with the Chern number.

In the $C=2$ region, using the parameter set $(\lambda_{SO},\mu)=(0.08,1.1)$ [Figs.~\ref{fig:p+ip}(a)], there are four energy bands crossing the Fermi energy at $k_x=0$ momentum, resulting in two pairs of Majorana zero modes at this momentum. In the $C=-4$ region, using the parameter set $(\lambda_{SO},\mu)=(0.22,-1.1)$ [Figs.~\ref{fig:p+ip}(c)], eight energy bands cross the Fermi energy, among which four energy bands cross at ${k_x} = 0$ and the other four cross at the momenta ${k_x} = \pm 0.86\pi$. In this case, two pairs of Majorana zero modes emerge at the momentum ${k_x} = 0$, and the other two pairs emerge at ${k_x} = \pm 0.86\pi$.

\begin{figure}
\includegraphics[width = 8cm]{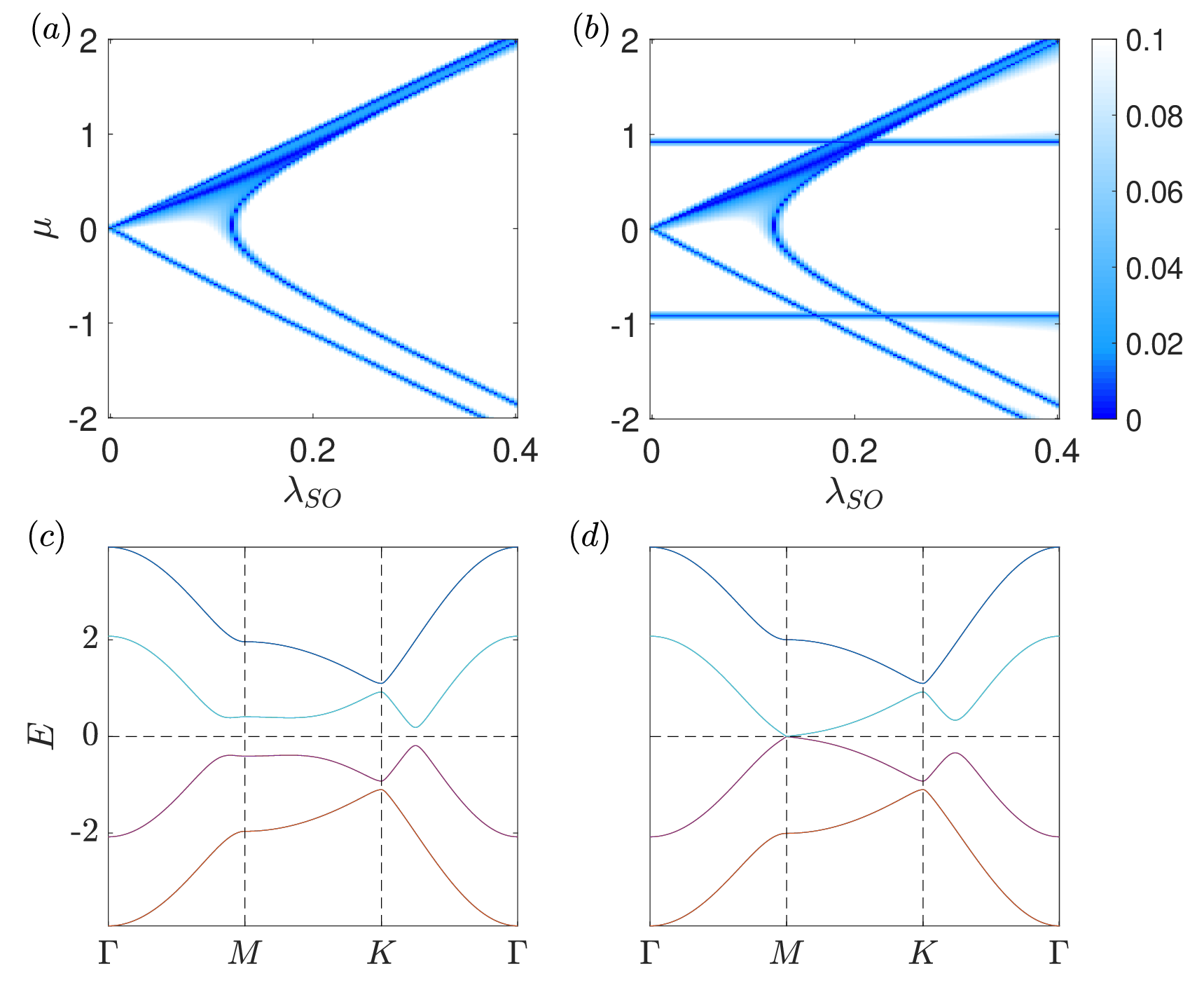}
\caption{\label{fig:gap} Electronic structures in the $d+id$ and $p+ip$ superconducting states. Panels (a) and (b) display intensity plots of the energy gap as functions of spin-orbit coupling strength $\lambda_{SO}$ and chemical potential $\mu$ for the $d+id$ pairing state and the $p+ip$ pairing state, respectively. Panels (c) and (d) illustrate the energy bands along highly symmetric lines of the first Brillouin zone for both states ($d+id$ and $p+ip$) without spin-orbital coupling and with a chemical potential of $\mu=0.9$. 
}
\end{figure}

\begin{figure*}
\centering
\includegraphics[width = 18cm]{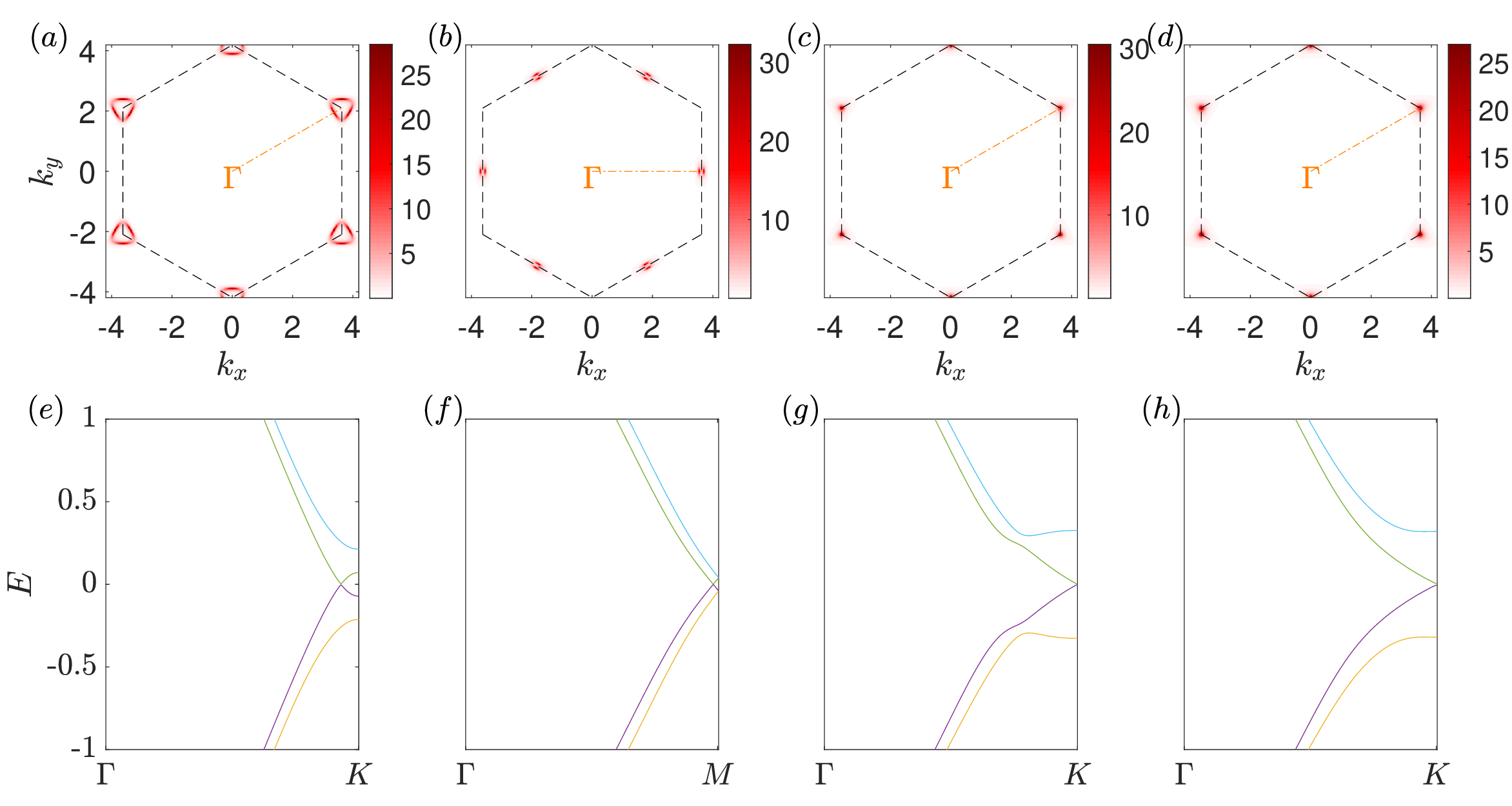}
\caption{\label{fig:SFpk}
Electronic structures along various phase transition points: Panels (a-d) display the intensity plots of zero-energy spectral functions for parameter sets $(\lambda_{SO},\mu)=(0.11,0.5)$, $(0.05,-0.9)$, $(0.21,-1.2)$, and $(0.28,-1.2)$, respectively. Panels (e-h) illustrate the corresponding energy bands along highly symmetric lines within the first Brillouin zone for these parameter sets.
}
\end{figure*}

In the $C=6$ region, using the parameter set $(\lambda_{SO},\mu)=(0.3,1.1)$ [Figs.~\ref{fig:p+ip}(e)], twelve energy bands cross the Fermi energy, among which four energy bands cross at ${k_x} = 0$ and the other eight cross at the momenta ${k_x} = \pm 0.9\pi\pm 0.01\pi$. In this case, two pairs of Majorana zero modes emerge at the momentum ${k_x} = 0$, and the other four pairs emerge at ${k_x} = \pm 0.9\pi\pm 0.01\pi$, resulting in six pairs of Majorana zero modes. The spectral functions at the system edge, presented in the right panels of Fig.~\ref{fig:p+ip}, also indicate that the edge modes are chiral, similar to the results of the $d+id$ pairing states.

\section{\label{sec:Discussion}Discussion}

At this stage, we aim to elucidate the origin of the exotic topological phase diagram in the chiral superconducting state. Generally, a single-band two-dimensional chiral superconductor is topologically nontrivial with a Chern number $C=\pm 1$. In the normal state, due to the presence of spin-orbit coupling and the two sub-lattices, there are four energy bands in the normal state. As a result, it is understandable that the $C=\pm 4$ topological phase initially emerges in the phase diagram.

 Topological phase transitions may occur when the parameters $\mu$ or $\lambda_{SO}$ change. Generally, along the topological transition lines, the energy gap tends to close. The intensity plots of the energy gap in the two pairing states, as functions of $\mu$ and $\lambda_{SO}$, are displayed in Figs.~\ref{fig:gap}(a) and ~\ref{fig:gap}(b), where the blue lines indicate the gap closing lines. In comparison to the phase diagrams shown in Fig.~\ref{fig:PD}, the energy gaps of the system indeed close along the phase transition lines.
The differences between the phase diagrams of the $d+id$ state and the $p+ip$ state can be effectively explained by examining the band structures of these two pairing states. Previous studies have extensively investigated the electronic structures in the superconducting state on the honeycomb lattice. The normal state energy band exhibits the VHS at the chemical potential around $\mu=1$. Near the VHS point, the energy gap in the $p+ip$ superconducting state is generally smaller than that in the $d+id$ superconducting state~\cite{Li2021}. Furthermore, the nodal or fully gapped states near the VHS point for graphene material have been previously investigated~\cite{PhysRevB.108.134514}.
It was indicated that for the $p+ip$ pairing symmetry, additional nodal points emerge near the $M$ point when the Fermi level is around the VHS point. In this work, we present numerical results of the energy bands along the highly symmetric lines near the VHS point $(\mu=0.9)$ and without spin-orbital coupling $(\lambda_{R}=\lambda_{SO}=0)$ for these two pairing states in Figs.~\ref{fig:gap}(c) and ~\ref{fig:gap}(d), respectively. As observed, in the $p+ip$ superconducting state, the energy gap closes and nodal points indeed emerge near the $M$ point. In contrast, in the $d+id$ superconducting state, the system remains fully gapped along the entire highly symmetrical lines. These results are consistent with the previous study~\cite{PhysRevB.108.134514}.
As a result, additional phase transition lines near the VHS point appear for the $p+ip$ pairing states, as illustrated in Fig.~\ref{fig:PD}.

We now attempt to elucidate the origin of the exotic topological phase diagram in the chiral superconducting state by examining the Dirac cones along phase transition lines in the phase diagrams. Within the topologically nontrivial phase, the bulk energy bands of the system are generally fully gapped, allowing the Chern number of the system to be defined. At the phase transition points, the energy gap typically closes, often accompanied by Dirac cones. The change in the Chern number should be related to the number of Dirac cones in the energy bands~\cite{srep28471}. 

Numerically, the position and number of Dirac cones can be explored by calculating the zero-energy spectral function. The phase diagram of the $d+id$ pairing state is relatively simple, and the main characteristics are involved in that of the $p+ip$ state. We focus on the phase transitions in the $p+ip$ pairing state. The intensity plots of the zero-energy spectral function at different phase transition points are presented in Figs. \ref{fig:SFpk}(a)-\ref{fig:SFpk}(d), respectively. The corresponding energy bands along the highly symmetric lines of the Brillouin zone are presented in Figs. \ref{fig:SFpk}(e)-\ref{fig:SFpk}(h).

As seen in Fig.~\ref{fig:PD}(b), the point $(\lambda_{SO},\mu)=(0.11,0.5)$ connects two topological phases of $C=4$ and $C=-2$. The numerical results of the electronic structure at this point are displayed in Figs.~\ref{fig:SFpk}(a) and \ref{fig:SFpk}(e). The gap is closed, and a Dirac cone emerges along the $\Gamma-K$ line. Due to the six-fold symmetry of the honeycomb lattice, there are actually six different Dirac cones in the first Brillouin zone, as seen in Fig.~\ref{fig:SFpk}(e). This explains why the Chern number decreases from 4 to -2 when crossing this point. Similarly, the point $(0.05,-0.9)$ connects the two phases with $C=4$ and $C=-2$. The corresponding numerical results of the electronic structure at this point are displayed in Figs.~\ref{fig:SFpk}(b) and \ref{fig:SFpk}(f). A Dirac cone emerges along the $\Gamma-M$ line. Also, in the first Brillouin zone, there are six different Dirac points due to the six-fold symmetry of the energy bands, consistent with the decrease in the Chern number when crossing this point.

\begin{figure}
\centering
\includegraphics[width = 8cm]{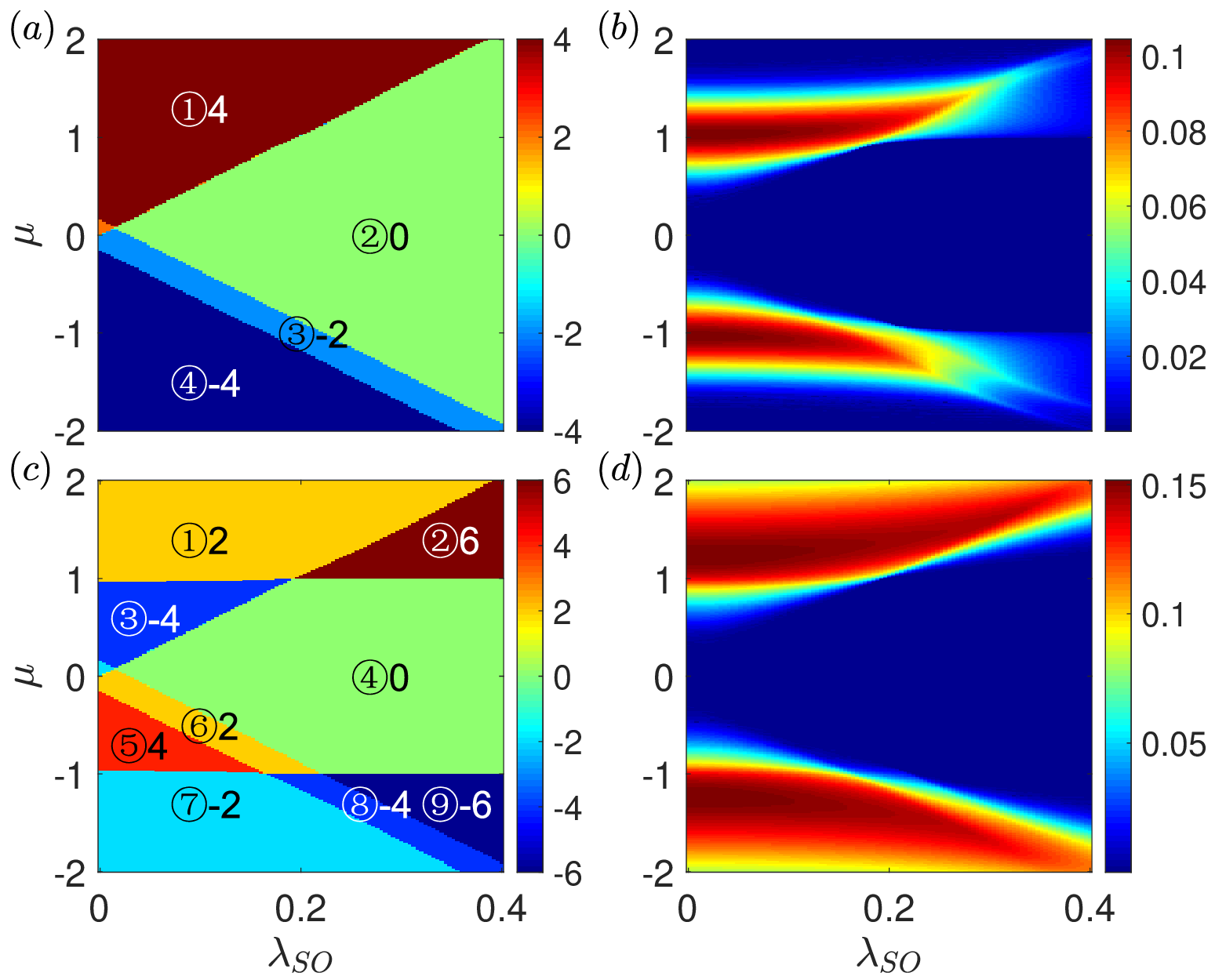}
\caption{\label{fig:PDself} 
Numerical results from self-consistent calculations: Panels (a) and (b) display the self-consistently obtained phase diagram and pairing magnitudes for the $d+id$ pairing state. Panels (c) and (d) show the corresponding results for the $p+ip$ pairing state.}
\end{figure}

We now discuss the energy bands at the points $(0.21,-1.2)$ and $(0.28,-1.2)$. Crossing these two points, the Chern number changes from $-2$ to $-4$ and from $-4$ to $-6$, respectively. As seen in Figs.~\ref{fig:SFpk}(g) and ~\ref{fig:SFpk}(h), the Dirac cones emerge at the $K$ points, namely, the vertices of the Brillouin zone. For the hexagonal Brillouin zone, among the six vertices, only two neighboring vertices belong to the first Brillouin zone. The other four vertices can be obtained by translating these two points through reciprocal lattice vectors. Therefore, at these two points, there are merely two different Dirac cones in the first Brillouin zone, resulting in a change of 2 in the Chern number when crossing these points.

It is necessary to discuss the robustness of the phase diagram against the superconducting pairing magnitude. In Fig.~\ref{fig:PD}, we consider a constant superconducting gap magnitude with $\Delta_0=0.2$. We have checked numerically that when the gap magnitude varies, the main features of the phase diagram remain qualitatively the same. Moreover, for an intrinsic superconductor, the superconducting magnitudes may strongly depend on the parameters. We also calculate the superconducting gap magnitudes self-consistently based on Eqs.(8) and (9) by considering an effective attractive interaction $V=2$. The self-consistently obtained phase diagram and the superconducting order parameters are plotted in Fig.~\ref{fig:PDself}. As seen, the order parameters indeed depend strongly on the parameters, while the main features of the phase diagram remain qualitatively similar. For the $d+id$ pairing state, the original $C=\pm 4$ topological phases emerge. The Chern number reduces as the spin-orbit coupling strength increases, and a large topologically trivial region appears in the phase diagram. For the $p+ip$ pairing states, additional phase transition lines appear near the VHS. The higher Chern number regions with $C=\pm 6$ emerge in the phase diagram. Our numerical results indicate that the exotic topological properties of the phase diagram are indeed quite robust, at least qualitatively.


\section{\label{sec:Summary}Summary}
In summary, we have explored the topological properties of the chiral superconducting state in the doped Kane-Mele model, considering two different pairing symmetries: the $d+id$ state and the $p+ip$ state. The phase diagram with different chemical potentials and spin-orbit coupling strengths is displayed, using the Chern number to describe various topological phases.

In the $d+id$ pairing state, higher Chern number states with $C=\pm 4$ emerge in the phase diagram. As the spin-orbit coupling strength increases, the Chern number decreases, and ultimately the system becomes topologically trivial. In the $p+ip$ state, the phase diagram at small chemical potentials is qualitatively similar to that in the $d+id$ state, while additional phase transition lines emerge in the heavily overdoped region with chemical potentials around $\mu=\pm 0.9$. Higher Chern number phases with $C=\pm 6$ emerge.
We also examine the number of edge states to verify the Chern number of different phases. All the exotic topological properties can be understood by studying the electronic structures at the phase transition points.

\begin{acknowledgments}
We thank Wei Chen and Wen Huang for useful discussions. This work was supported by the NSFC (Grant No.12074130), the Natural Science Foundation of Guangdong Province (Grant No. 2021A1515012340), the Key-Area Research and
Development Program of Guangdong Province (Grant No.2019B030330001), and the CRF of Hong Kong (C6009-20G).
\end{acknowledgments}


\providecommand{\noopsort}[1]{}\providecommand{\singleletter}[1]{#1}%

\end{document}